\documentclass[letterpaper, 10 pt, conference]{ieeeconf}  

\IEEEoverridecommandlockouts                              

\usepackage{graphics} 
\usepackage{epsfig} 
\usepackage{mathptmx} 
\usepackage{times} 
\usepackage{amsmath} 
\usepackage{amssymb}  
\usepackage{subcaption}
\usepackage{booktabs}
\usepackage{blkarray}
\usepackage{xfrac}
\usepackage{xcolor}
\usepackage{wrapfig}
\usepackage[linesnumbered,lined,ruled,commentsnumbered]{algorithm2e}

\DeclareMathOperator*{\argmin}{arg\,min}
\DeclareMathOperator*{\argmax}{arg\,max}

\usepackage{mathtools}

\usepackage{scalerel,stackengine}
\stackMath
\newcommand\reallywidehat[1]{%
	\savestack{\tmpbox}{\stretchto{%
			\scaleto{%
				\scalerel*[\widthof{\ensuremath{#1}}]{\kern-.6pt\bigwedge\kern-.6pt}%
				{\rule[-\textheight/2]{1ex}{\textheight}}
			}{\textheight}%
		}{0.5ex}}%
	\stackon[1pt]{#1}{\tmpbox}%
}

\DeclareMathOperator{\Tr}{Tr}
\DeclareMathOperator{\CDF}{CDF}

\title{\LARGE \bf
	Coordinated Control of UAVs for Human-Centered Active Sensing of Wildfires
}

\author{Esmaeil Seraj$^{1}$ and Matthew Gombolay$^{1}$
	\thanks{$^{1}$Institute for Robotics and Intelligent Machines, Georgia Institute of Technology, Atlanta, GA 30332, USA
		{\tt\small eseraj3@gatech.edu, matthew.gombolay@cc.gatech.edu}}%
}

\begin{document}

	\maketitle
	\thispagestyle{empty}
	\pagestyle{empty}

	\begin{abstract}
		Fighting wildfires is a precarious task, imperiling the lives of engaging firefighters and those who reside in the fire's path. Firefighters need online and dynamic observation of the firefront to anticipate a wildfire's unknown characteristics, such as size, scale, and propagation velocity, and to plan accordingly. In this paper, we propose a distributed control framework to coordinate a team of unmanned aerial vehicles (UAVs) for a human-centered active sensing of wildfires. We develop a dual-criterion objective function based on Kalman uncertainty residual propagation and weighted multi-agent consensus protocol, which enables the UAVs to actively infer the wildfire dynamics and parameters, track and monitor the fire transition, and safely manage human firefighters on the ground using acquired information. We evaluate our approach relative to prior work, showing significant improvements by reducing the environment’s cumulative uncertainty residual by more than $ 10^2 $ and $ 10^5 $ times in firefront coverage performance to support human-robot teaming for firefighting. We also demonstrate our method on physical robots in a mock firefighting exercise.
	\end{abstract}

	\section{Introduction}
	\label{sec:introduction}
\begin{figure}[t!]
		\centering
		\includegraphics[width=0.8\columnwidth]{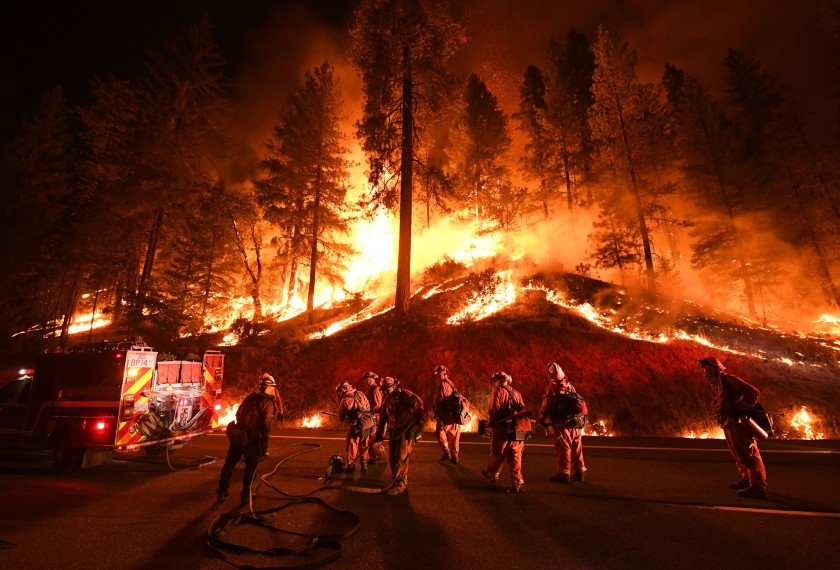}
		\caption{Firefighters trying to control a back burn as the Carr fire spreads toward Douglas City. Courtesy of the Los Angeles Times (May 2019).}
		\label{fig:CaliWildFire}
		\vspace{-0.5cm}
	\end{figure}

	\noindent
	Fighting wildfires is a dangerous task and requires accurate online information regarding firefront location, size and scale, shape, and propagation velocity~\cite{sujit2007cooperative}. Firefighters may lose their lives as a consequence of inaccurately anticipating information either due to inherent stochasticity in fire behavior or low-quality and unusable information provided, such as low-resolution satellite images~\cite{casbeer2006cooperative, kudoh2003two}. Firefighters need frequent, high-quality images to monitor the fire propagation and plan accordingly (Fig.~\ref{fig:CaliWildFire}). Due to recent advances in aerial robotic technology, UAVs have been proposed as a solution to overcoming the challenges of needing real-time information in fighting fires~\cite{ollero2006unmanned}.
	
	In \cite{sujit2007cooperative}, a cooperative approach is proposed to detect local fire areas in a wildfire using two groups of detector and service UAV agents. In \cite{merino2010automatic}, utility of visual and infrared cameras on heterogeneous UAVs with hovering capabilities is investigated to monitor the evolution of the firefront shape. In \cite{ghamry2016cooperative}, a leader-follower-based distributed framework is proposed for a team of UAVs to evenly distribute and track an elliptical fire perimeter. In \cite{pham2017distributed}, a heat-intensity-based distributed control framework is designed for a team of UAVs to be capable of closely monitoring a wildfire in open space. More recently, both model-based (i.e., Kalman estimation) and learning-based (deep convolutional neural network) methods have been used for cooperative prediction and tracking of the firefront shape~\cite{zhou2019ensemble, lin2018kalman, kumar2011cooperative}. Additionally, other learning-based approaches, such as reinforcement learning (RL), have also been applied to this problem to enable collaborative monitoring of wildfires \cite{haksar2018distributed, julian2019distributed}, which can be enabled for online data processing through NN pruning approaches~\cite{karimzadeh2019hardware}.
	
	Many of the aforementioned studies require an accurate function for fire-shape~\cite{kumar2011cooperative} to work, assuming an enlarging elliptical perimeter to be monitored by UAVs~\cite{sujit2007cooperative, casbeer2006cooperative, ghamry2016cooperative}. Supposing a shape model for large-scale wildfires is not realistic and thus not accurate~\cite{morvan2011physical}. While vision-based approaches are still struggling with fire smoke elimination and image stabilization problems~\cite{yuan2015survey}, other approaches, such as~\cite{pham2017distributed}, require a heat-intensity model and are dependent on an accurate estimation of maximum heat-intensity within the entire fire-map. Additionally, ~\cite{seraj2019safe} notes that RL and learning-based methods are prone to major drawbacks, such as scalability and domain shift problems as well as lack of formal guarantees on boundedness of errors, which are significant for safety-critical applications, such as firefighting.
	
	There is a clear absence of human-centric approaches in the literature for UAV teams for active sensing of wildfire and fire monitoring. This is mainly because a majority of previous studies are solely focused on autonomous fire detection and surveilling a large burning area by drones rather than focusing on local human-defined areas of priority (areas of firefighter activity) and serving firefighters. In this study, we seek a better control strategy, toward a human-centered robot coordination, through better perception and accurate local situational awareness. We overcome key limitations in prior work by developing an algorithmic framework to provide a model-predictive mechanism that enables firefighters on the ground to receive online, high-quality information regarding their time-varying proximity to a fire. 
	
	In our approach, we explicitly estimate the latent fire propagation dynamics and parameters via an adaptive extended Kalman filter (AEKF) predictor and the simplified FARSITE wildfire propagation model~\cite{finney1998farsite} to account for firefighter's safety and provide them with online information regarding propagating firefronts. This model allows us to develop straightforward distributed control adapted from vehicle routing literature~\cite{toth2002vehicle} to enable track-based fire coverage. Moreover, a mathematical observation model through which UAV sensors observe fire is derived to map from state space to observation space. The calculated models are then used in combination to derive a dual-criteria objective function in order to control a fleet of UAVs. The proposed dual-criteria objective is an ad hoc, well-suited function to the wildfire monitoring task, which minimizes environment's uncertainty on local, human-centered areas (first criteria) and maximizes coverage through ensemble-level formation control of the robot network (second criteria).
	
	We empirically evaluate our approach against simulated wildfires alongside contemporary approaches for UAV coverage~\cite{pham2017distributed} as well as against a reinforcement learning baseline, demonstrating a promising utility of our approach. Our proposed coordinated controller is capable of reducing the cumulative uncertainty residual of the fire environment by more than $ 10^2 $ and $ 10^5 $ times in firefront coverage performance to support human-robot teaming for firefighting. We also assess the feasibility of our method through implementation on physical robots in a mock firefighting scenario.

	\section{Preliminaries}
	\label{sec:preliminaries}
	\noindent In this section, we first introduce the simplified FARSITE wildfire propagation mathematical model and calculate the fire dynamics. We then review the fundamentals of AEKF.
	
	\subsection{Fire Area Simulator (FARSITE): Fire Propagation Model}
	\label{subsec:simplifiedfarsite}
	\noindent The Fire Area Simulator (FARSITE) wildfire propagation model was first introduced by Finney et al \cite{finney1998farsite}, which is now widely used by the USDI National Park Service, USDA Forest Service, and other land management agencies at the federal and state levels. The model has been utilized to simulate the spread of a wildfire, factoring in heterogeneous conditions of terrain, fuels, and weather and their influence on fire dynamics. The full FARSITE model includes complex equations, and precise model implementation requires a significant amount of geographical and physical information on terrain, fuels, and weather. As such, researchers tend to modify the model by considering few simplifying assumptions~\cite{pham2017distributed}. The wildfire propagation dynamics using a simplified FARSITE model are shown in Eq.~\ref{eq:firedynamics1} and \ref{eq:qdotdefinition}.
	\par\nobreak{\parskip0pt \small \noindent 
		\begin{align}
			q_t^i &= q_{t-1}^i + \dot{q}_{t-1}^i\delta t \label{eq:firedynamics1}\\
			\dot{q}_{t}^i &= \frac{d}{dt}\left(q_t^i\right)\label{eq:qdotdefinition}
	\end{align}}In the above equations, $ q_t^i $ indicates the location of firefront $ i $ at time $ t $, and $ \dot{q}_{t}^i $ is its growth rate (i.e., propagation velocity). $ \dot{q}_{t} $ is a function of fire spread rate ($ R_t $), wind speed ($ U_t $), and wind azimuth ($ \theta_t $), which are available to our system through weather forecasting equipment. By ignoring the superscript $ i $ in Eq.~\ref{eq:firedynamics1} and~\ref{eq:qdotdefinition} and without losing generality, $ \dot{q}_{t} $ can be estimated for each propagating firefront by Eq.~\ref{eq:qdot1}-\ref{eq:qdot2}, where $ \dot{q}_{t}^x $ and $ \dot{q}_{t}^x $ are first-order firefront dynamics for \textit{X} and \textit{Y} axes~\cite{finney1998farsite}
	\par\nobreak{\parskip0pt \small \noindent 
		\begin{align}
			\dot{q}_{t}^x &= C(R_t, U_t)\sin(\theta_t) \label{eq:qdot1} \\
			\dot{q}_{t}^y &= C(R_t, U_t)\cos(\theta_t)     \label{eq:qdot2}
	\end{align}}where \nobreak{\parskip0pt \small \noindent$ C(R_t, U_t) = R_t\left(1-\frac{LB(U_t)}{LB(U_t) + \sqrt{GB(U_t)}}\right) $} in which \nobreak{\parskip0pt \small \noindent$ LB(U_t) = 0.936e^{0.256U_t} + 0.461e^{-0.154U_t} - 0.397 $} and \nobreak{\parskip0pt \small \noindent$ GB(U_t) = LB(U_t)^2 - 1 $}.

	\subsection{Adaptive Extended Kalman Filter (AEKF)}
	\label{subsec:EKF}
	\noindent We utilize an adaptive extended Kalman filter (AEKF) to leverage the mathematical fire propagation model and the observation model of a flying drone with respect to a dynamic object on the ground to actively sense the fire-spots, infer wildfire dynamics and parameters, and propagate all sources of measurement uncertainty. In a conventional extended Kalman filter (EKF), process and observation noise covariances $ Q_t $ and $ \Gamma_t $ are often chosen as constant matrices based on the state-transition model and sensor accuracy and do not receive updates. However, this selection process is highly sensitive to user experience and can be extremely inaccurate. We leverage AEKF~\cite{akhlaghi2017adaptive}, which introduces \textit{innovation} and \textit{residual}-based updates for process and observation noise covariances, as shown in Eq.~\ref{eq:adaptiveQ}-\ref{eq:adaptiver}, where $ \alpha $ is a forgetting factor and $ d_t $ is the measurement innovation and is defined as the difference between the actual measurement and its predicted value. Moreover, $ K_t $ is the Kalman gain and $ H_t $ is the observation Jacobian matrix.
	\par\nobreak{\parskip0pt \small \noindent 
		\begin{align}
			Q_t &= \alpha Q_{t-1} + (1-\alpha)\left(K_td_td_t^TK_t^T\right) \label{eq:adaptiveQ} \\
			\Gamma_t &= \alpha \Gamma_{t-1} + (1-\alpha)\left(\tilde{y}_t\tilde{y}_t^T + H_tP_{t|t-1}H_t^T\right) \label{eq:adaptiver}
	\end{align}}These adaptive updates remove the assumption of constant covariances $Q_t$ and $\Gamma_t$ and enable even more accurate predictions over time as the Kalman filter leverages its observations to improve the predicted covariance matrix $ P_{t|t-1} $~\cite{akhlaghi2017adaptive}.
	
	\section{Problem Statement and Algorithmic Overview}
	\label{sec:problemstatement}
	\noindent The focus of our study includes two important aspects of wildfire fighting: (1) providing high-quality information on firefront status while accounting for physical and methodological errors and (2) human-centered coverage and tracking of wildfire to account for firefighter safety. Accordingly, we define high-quality information as high-resolution and online images of areas prioritized by humans. We take advantage of the estimated uncertainty of the environment to achieve these objectives. Through a unified error propagation system, not only can the physical and methodological uncertainties be leveraged to manage the human teams and account for their safety, they can also be used to manage the UAV team, both in node-level and ensemble-level dynamics. An AEKF is a proper candidate for the uncertainty propagation system here since it can accumulate physical and methodological errors and generate a cumulative error map through the calculated probability distribution and predicted covariance matrix. 
	
	Accordingly, UAVs initially calculate two uncertainty maps: (1) a firefront uncertainty map (Section~\ref{subsubsec:firefrontUncertaintyMap}) and (2) a human uncertainty map (Section~\ref{subsubsec:humanUncertaintyMap}). The latter is generated through a bimodal distribution of human locations as received by GPS devices while the first map is created by the AEKF's online inference of firefront locations and error propagation (Section~\ref{subsec:OnlineInferenceOfWildfireDynamics}). Through the combination of these two error-maps, we obtain our first node-level controller (uncertainty-based controller, Section~\ref{subsec:generatinguncertaintymap&predictingfirefrontcharacteristics}). We also incorporate an ensemble-level (formation) controller to encourage the UAV team to maintain a formation consensus for maximizing the coverage (Section~\ref{subsec:FormationController}). The two controllers coordinate to generate a virtual position for each UAV which is then fed to a path planning controller to generate the force required to move the UAV to the determined position based on UAVs flight dynamics (Section~\ref{sec:controllerdesign}).
	
	\section{Method}
	\label{sec:method}
	\noindent Algorithm~\ref{alg:method} depicts an overview of the proposed human-centered coordinated control procedure for monitoring wildfires. Upon receiving a request, UAVs travel to the human-defined areas of interest (i.e., rendezvous point, line 1 in Algorithm~\ref{alg:method}). On arrival, UAVs sense the firefront by extrapolating fire-spots $ q_t $ and generate a general uncertainty map $ \mathcal{U}_t^{tot} $ by fusing AEKF error propagation and areas of human activity using GPS data (line 3-5 in Algorithm~\ref{alg:method}). Afterwards, a combination of an uncertainty-based optimization and a graph-based weighted consensus protocol forms our new dual-criteria objective function $ \mathcal{H} $ for a set of $ N $ UAVs (line 6 in Algorithm~\ref{alg:method}). This objective function is then embedded as our coordinated control system to move UAVs to highly uncertain areas on the generated error map to minimize the associated error while encouraging drones to maintain a distributed formation to increase the team efficiency in field coverage (lines 7-8 in Algorithm~\ref{alg:method}). Meanwhile, AEKF is also used to infer the firefront characteristics, such as spatial distribution $ \hat{q}_t $, propagation velocity $ \dot{q}_t $, and direction , in order to calculate an individualized temporal safety index (SI) (Eq.~\ref{eq:safetyindex}) for firefighters.
	\par\nobreak{\parskip0pt \footnotesize \noindent
		\begin{equation}
			\label{eq:dualobjfunc}
			\min\mathcal{H} = \underset{q_t^i, p_t^d}{\arg\max}\left(\int_{i\in Q}\mathcal{U}_t^{tot}\left(q_t^i\right)dq - \int_{d\in N}\mathcal{E}_{dj}\left(\left\|p_t^d - p_t^{j/d}\right\|\right)dp\right)
	\end{equation}}where $ p_t^d $ represents UAV positions and $ Q $ is the entire fire-map. To generate the required control inputs for UAVs to move, we calculate the negative derivative of the objective function with respect to the location of drones at time $ t $. 
	
	The following sections are dedicated to discussing and formulating the two modules of the dual-criteria objective function in Eq.~\ref{eq:dualobjfunc}, as well as elaborating on the uncertainty map generation process.
	\begin{algorithm}
		\footnotesize
		\SetKwData{NumUAV}{NumUAV}\SetKwData{ErrorMap}{ErrorMap}\SetKwData{MissionDuration}{MissionDuration}
		\SetKwFunction{GPS}{GPS}\SetKwFunction{Sense}{Sense}\SetKwFunction{Move}{Move}
		\SetKwInOut{Input}{input}\SetKwInOut{Output}{output}
		\Input{Obtain the rendezvous area $ p_r $, fire-map $ Q_t $, human GPS data $ p_t^h $, UAV positions $ p_t^d $, and fire propagation model $ \mathcal{M} $}
		\BlankLine
		
		Move to rendezvous area: $ p_{t}^d \leftarrow \Move( p_r, p_{t-1}^d ) $
		
		\While{MissionDuration}{
			
			Generate firefront uncertainty map: $ \mathcal{U}_t^{f}, q_t \leftarrow \Sense(Q_t) $
			
			Generate human uncertainty map: $ \mathcal{U}_t^{h} \leftarrow \GPS(q_h) $
			
			Combine uncertainty maps: $ \mathcal{U}_t^{tot} = \mathcal{U}_t^{f} + \mathcal{U}_t^{h} $
			
			Minimize the dual-criteria objective function:  \scriptsize{$$ \min\mathcal{H} = \underset{q_t^i, p_t^d}{\arg\max}\left(\int_{i\in Q}\mathcal{U}_t^{tot}\left(q_t^i\right)dq - \int_{d\in N}\mathcal{E}_{dj}\left(\left\|p_t^d - p_t^{j/d}\right\|\right)dp\right) $$}

			Calculate the overall control inputs and determine new virtual positions for UAVs: $$ p_{t+\delta t}^n=p_t^v-\left(u_d^{ucc} - u_d^{fcc}\right)\delta t $$

			Move to the new desired position: $ p_{t+\delta t}^d \leftarrow \Move(p_{t+\delta t}^n, p_{t}^d ) $
			
		}
		\textbf{def} ~\Move{$ p_g^d, p_t^d $}\textbf{:} ~~~// $ p_g^d $ is the goal position for drone $ d $
		
		~\hspace{0.75cm}$ u_{d,t} = \sum_{i}u_{d,t}^{att_i}(p_g^d, p_t^d)+u_{d,t}^{rep_i}(p_g^d, p_t^d), ~~\forall i \in F_t $
		
		~\hspace{0.75cm}$ p_{t+\delta t}^d= p_t^d+u_{d,t}\delta t $

		\textbf{def} ~\Sense{$ Q_t $}\textbf{:} ~~~// $ \mathcal{O}_t $ is the UAV observation model
		
		
		~\hspace{0.75cm}$ \hat{q}_t = \argmax_{q_t} \rho\left(q_{t|t-1}, p_{t-1}, \mathcal{M}_{t-1}, \mathcal{O}_{t-1}\right) $  
		
		\caption{Stages of the proposed human-centered coordinated control for a team of UAVs.}
		
		\label{alg:method}
		
	\end{algorithm}
	
	\subsection{Criteria 1: Uncertainty-based Controller}
	\label{subsec:generatinguncertaintymap&predictingfirefrontcharacteristics}
\begin{figure}[t!]
		\centering
		\includegraphics[width=\columnwidth]{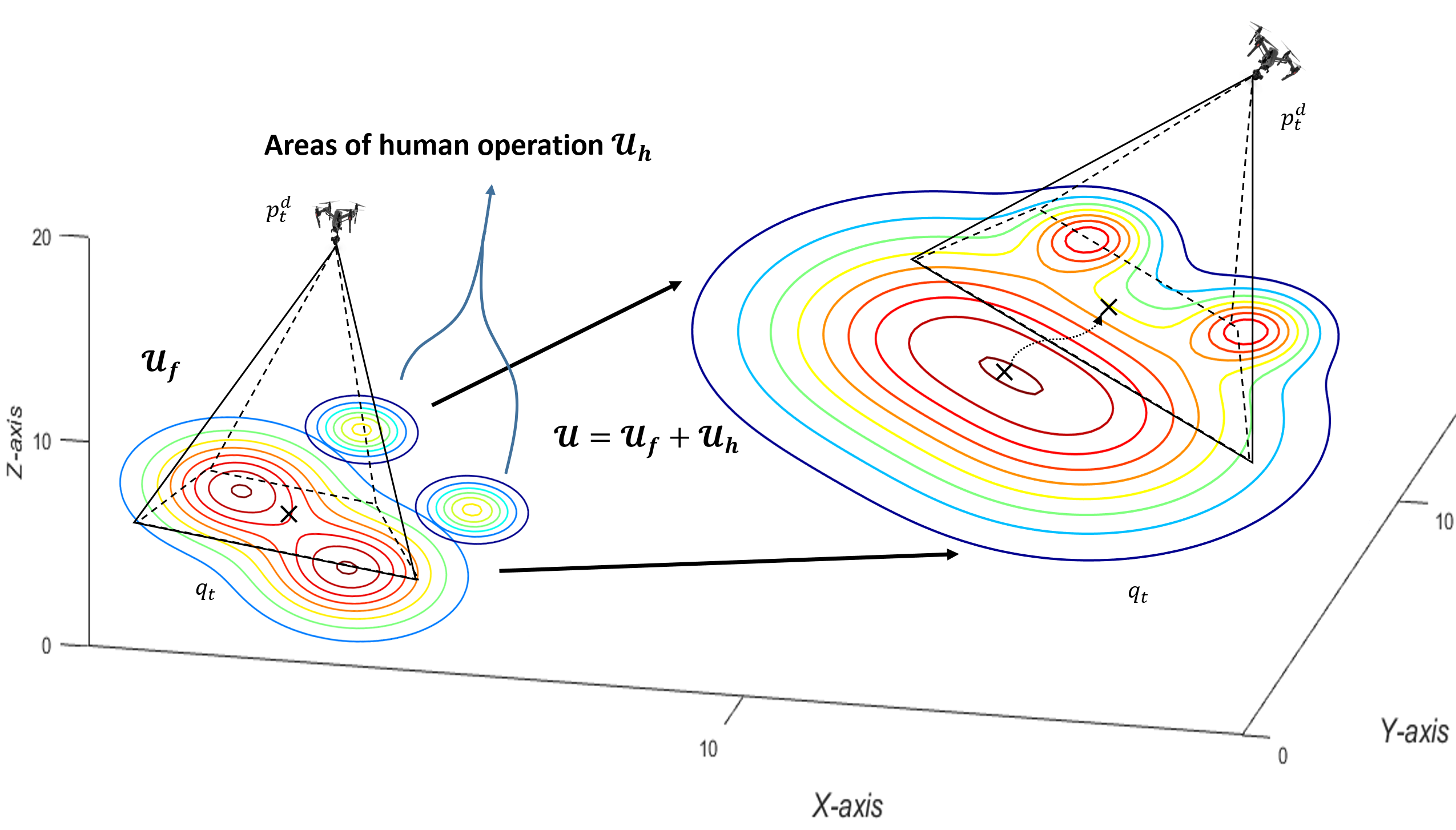}
		\caption{Fusing AEKF measurement residual and human GPS data to generate an uncertainty map. UAVs focus on areas of human activity while closely monitoring the fire propagation.}
		\label{fig:EKF_ErrorMap}
		\vspace{-0.5cm}
	\end{figure}

	\noindent We design our coverage and tracking controller to minimize the uncertainty of the firefront locations over time, while focusing on the areas of human operation. To this end, we generate two uncertainty maps for (1) the propagating fire locations $ \mathcal{U}_f $ and (2) the areas of human activity $ \mathcal{U}_h $. Eventually, we fuse these error maps together by linearly summing up the respective estimated uncertainty values of each point to obtain the general uncertainty map $ \mathcal{U}_t^{tot} $ at time $ t $. The uncertainty-based controller's objective is to minimize the overall uncertainty residual in $ \mathcal{U}_t^{tot} $. We present the details of calculating $ \mathcal{U}_f $ and $ \mathcal{U}_h $ in the following sections. Fig.~\ref{fig:EKF_ErrorMap} demonstrates the formation of the uncertainty map and the foundation of our human-centered controller.
	
	\subsubsection{Firefront Uncertainty Map}
	\label{subsubsec:firefrontUncertaintyMap}
	We leverage AEKF to estimate a probability distribution of the fire-spot locations and compute a measurement covariance for each point through linear error propagation techniques. Considering $ q_{t-1} $ as the location of firefronts at current time and $ p_{t-1}^d $ as the UAV coordinates, a firefront location $ \hat{q}_t $ one step forward in time is desired, given the current firefront distribution ($ q_{t-1} $), fire propagation model with current parameters ($ \mathcal{M}_{t-1} $), and UAV observation model of the field ($ \mathcal{O}_{t-1} $) as in Eq.~\ref{eq:predproblem}
	\par\nobreak{\parskip0pt \small \noindent 
		\begin{equation}
			\label{eq:predproblem}
			\hat{q}_t = \argmax_{q_t} \rho\left(q_{t-1}, p_{t-1}, \mathcal{M}_{t-1}, \mathcal{O}_{t-1}, q_t\right)
	\end{equation}}In AEKF, the uncertainty of the firefront locations over time is measured as the state covariance $ P_{t|t} $ at time $ t $. It has been shown previously (see~\cite{sim2005stable}) that minimizing the state covariance corresponds to maximizing the covariance residual $ S_t $ in Eq.~\ref{eq:covres}
	\par\nobreak{\parskip0pt \small \noindent 
		\begin{equation}
			\label{eq:covres}
			S_t = H_t P_{t|t-1}H_t^T + \Gamma_t
	\end{equation}}where $ P_{t|t-1} $ is the predicted covariance, $ F_t $ and $ H_t $ are process and observation model Jacobians, and $ Q_t $ and $ \Gamma_t $ are the corresponding noise covariances and can be calculated as $ P_{t|t-1} = F_tP_{t-1|t-1}F_t^T + Q_t $. According to Eq.~\ref{eq:covres}, by setting $ p_{t|t-1} $ to identity, we see that a maximally informative position for drones is the one that minimizes the $ H_tH_t^T $, or in other words, the closest possible position where dynamic observations change rapidly~\cite{sim2005stable}. As such, we generate an uncertainty map which is reflective of the wildfire dynamics where a measurement residual can be calculated for each point $ q_t $ by summing up the estimated covariance residual matrix $ S_t $ and set our objective to maximize $ S_t $. Accordingly, we derive our new objective function $ \mathcal{H} $ as in Eq.~\ref{eq:newobjfunc2}, where $ \Tr(.) $ represents the trace operation, and $ fov_t^d $ is the field of view (FOV) of drone $ d $ at time $ t $.
	\par\nobreak{\parskip0pt \small \noindent
		\begin{align}
			\min\mathcal{U}_1 &= \underset{q_t^i\in fov_t^d}{\arg\max}\left(\Tr\left(S_t\right)\right) = \underset{q_t^i\in fov_t^d}{\arg\max}\left(\Tr\left(H_t P_{t|t-1}H_t^T + \Gamma_t\right)\right) \nonumber \\
			&= \underset{q_t^i\in fov_t^d}{\arg\max}\left(\Tr\left(H_t \left(F_tP_{t-1|t-1}F_t^T + Q_t\right)H_t^T + \Gamma_t\right)\right) \label{eq:newobjfunc2}
	\end{align}}Details of online inference of the parameters in the above equation are presented in Section~\ref{subsec:OnlineInferenceOfWildfireDynamics}.
	
	\subsubsection{Human Uncertainty Map}
	\label{subsubsec:humanUncertaintyMap}
	While hovering around the highly uncertain areas to provide firefighters with online information regarding the firefront, UAVs are required to focus on their human collaborators on the ground and take their safety into account by putting additional concentration on the areas of human operation. Accordingly, UAVs receive human positions $ p_t^h $ (i.e., through GPS devices) at time $ t $ as planar coordinates $ \mu_{t, h}^x $ and $ \mu_{t, h}^y $ and generate a bimodal Gaussian distribution for each human $ h $ to account for both error in GPS information as well as the mobility of the humans. By assuming independence, a joint PDF can be calculated as our human safety objective as in Eq.~\ref{eq:HSCobjfunc}
	\par\nobreak{\parskip0pt \small \noindent 
		\begin{equation}
			\label{eq:HSCobjfunc}
			\mathcal{U}_2 = \argmax_{q_t\in v_h}\prod_{i\in v_h}P_{ih}\left\{ q_t^i - p_t^h \geq r_s \right\} \geq P_s
	\end{equation}}where $ v_j $ represents the points in a safe circular vicinity of human $ h $ with radius $ r_s $ and $ P_s $ is a predefined safety threshold for probability. $ P_{ih} $ is a cumulative distribution function (CDF) with respect to each human location $ p_t^h $ and all approaching firefronts $ q_t^i $. To calculate the $ P_{ih} $, we leverage the estimated fire-spot locations $ q_t $ alongside inferred fire parameters (i.e., $ \hat{R}_t $, $ \hat{U}_t $, and $ \hat{\theta}_t $ ) by AEKF to calculate a CDF for each human at location $ p_t^h $ and all approaching firefronts $ q_t^i $. We then integrate the resulting CDF to be greater than the safe-distance. A Probability $ P_{ih} $ is then calculated for each individual as in Eq.~\ref{eq:CDFprob}
	\par\nobreak{\parskip0pt \footnotesize \noindent 
		\begin{equation}
			\label{eq:CDFprob}
			P_{ih} =  \int_{\tau=l_s}^{\infty}\mathcal{N}\left(\mu_{q_t^i} - p_t^h, \sigma_{t, i}^2\right)d\tau =  1-\CDF\left(r_s|\mu_{q_t^i} - p_t^h, \sigma_{t, i}^2\right)
	\end{equation}}where $ \mu_{q_t^i}$ and $ \sigma_{t, i}^2 $ are calculated by AEKF (see Section~\ref{subsec:OnlineInferenceOfWildfireDynamics}). Eq.~\ref{eq:HSCobjfunc} is leveraged later in Section~\ref{subsec:safetyIndex} to compute an individualized safety index for each firefighter.
	
	\subsection{Criteria 2: Weighted Multi-agent Consensus Protocol}
	\label{subsec:HSC}
	\noindent To ensure that the combination of the above objective functions results in local actions leading up to appropriate global performance, we enforce an extra control term in such a way that the UAVs also act on other easily measurable information, such as the relative displacements to neighboring drones. This is specifically important to disperse UAVs, while preserving the connectedness of the network, from converging to an extreme minima (i.e., a highly uncertain point). Accordingly, we leverage the weighted consensus protocol~\cite{cortes2017coordinated} as in Eq.~\ref{eq:disperrorobj} for a set of $ N $ UAVs with the objective of minimizing the total displacement error $ \mathcal{E}_{ij} $ while preserving a distance of at least $ \delta $ between all UAVs
	\par\nobreak{\parskip0pt \small \noindent
		\begin{align}
			\min\mathcal{E} &= \argmin_{p_t}\mathcal{E}_{ij}\left(\left\|p_t^i - p_t^j\right\|\right) \label{eq:disperrorobj}\\
			&=\argmin_{p_t}\sum_{i=1}^{N}\sum_{j\in V_i}\frac{1}{2(\Delta-\delta)}\left(\frac{\left\|p_t^i - p_t^j\right\|-\delta}{\Delta-\left\|p_t^i - p_t^j\right\|}\right)^2 \label{eq:disperrorobj1}
	\end{align}}where $ j\in V_i $ represents $ j-$th UAV within the communication range $ \Delta $ of UAV $ i $. In this way, a negative force will be generated to move UAVs apart from or closer to each other if they are getting closer than $ \delta $ or farther than $ \Delta $ (i.e., the UAV network will become disconnected). Note that $ \delta $ should be set high enough so that the UAV team can spread effectively.
	
	\section{Online Inference of Wildfire Dynamics}
	\label{subsec:OnlineInferenceOfWildfireDynamics}
	\noindent The joint probability density function in Eq.~\ref{eq:predproblem} is calculated through AEKF estimator. Using the aforementioned notations, the AEKF state transition and observation equations can now be stated as in Eq.~\ref{eq:probform1} and \ref{eq:probform3}
	\par\nobreak{\parskip0pt \small \noindent 
		\begin{align}
			\label{eq:probform1}
			q_t = & f_{t-1}\left(q_{t-1}, p_{t-1}, R_{t-1}, U_{t-1}, \theta_{t-1}\right) + \omega_t \\
			\label{eq:probform3}
			\hat{q}_t = & h_{t}\left(q_t, p_{t-1}\right) + \nu_t
	\end{align}}where $ p_{t-1} $ is the physical location of UAV. We reform the state transition Eq. in~\ref{eq:probform1} to account for all state variables in $ \Theta_t = \left[ q_t^x, q_t^y, p_t^x, p_t^y, p_t^z, R_t, U_t, \theta_t \right]^T $ as in Eq.~\ref{eq:probforrm2}
	\par\nobreak{\parskip0pt \small \noindent 
		\begin{equation}
			\label{eq:probforrm2}
			\Bigg[ \Theta_t \Bigg]_{8\times1} =
			\Bigg[ \left. \frac{\partial f}{\partial \Theta_i}\right|_{\hat{\Theta}_{t-1|t-1}} \Bigg]_{8\times8}
			\Bigg[ \Theta_{t-1} \Bigg]_{8\times1} + \omega_t, ~\forall i \in \Theta
	\end{equation}}where the process noise $  \omega_t $. Therefore, we form the state transition Jacobian matrices $ F_t $ as in Eq.~\ref{eq:statetransitionJacob}, including partial derivatives of wildfire propagation dynamics in Eq.~\ref{eq:firedynamics1} and~\ref{eq:qdotdefinition} with respect to all variables in state vector $ \Theta_t $.
	\par\nobreak{\parskip0pt \small \noindent 
		\begin{equation}
			\label{eq:statetransitionJacob}
			\left. \frac{\partial f}{\partial \Theta_i}\right|_{\hat{\Phi}_{t'}} =
			\begin{blockarray}{ccccccc}
				& q_{t^\prime}^x & q_{t^\prime}^y & p_{t^\prime}^{(3)} & R_{t^\prime} & U_{t^\prime} & \theta_{t^\prime} \\
				\begin{block}{c(cccccc)}
					q_t^x & 1 & 0 & 0^{(3)} & \frac{\partial q_{t}^x}{\partial R_{t^\prime}}  & \frac{\partial q_{t}^x}{\partial U_{t^\prime}} & \frac{\partial q_{t}^x}{\partial \theta_{t^\prime}} \\
					q_t^y & 0 & 1 & 0^{(3)} & \frac{\partial q_{t}^y}{\partial R_{t^\prime}}  & \frac{\partial q_{t}^y}{\partial U_{t^\prime}} & \frac{\partial q_{t}^y}{\partial \theta_{t^\prime}} \\
					p_{t}^{(3)} & 0 & 0 & 0^{(3)} & 0  & 0 & 0  \\
					R_t & 0 & 0 & 0^{(3)} & 1  & 0 & 0  \\
					U_t & 0 & 0 & 0^{(3)} & 0  & 1 & 0  \\
					\theta_t & 0 & 0 & 0^{(3)} & 0  & 0 & 1  \\
				\end{block}
			\end{blockarray}
	\end{equation}}where $ t' = t - 1 $ and superscript (3) represent number of column and row repetitions for all $ \left[p_t^x, p_t^y, p_t^z\right] $. We note that the parameters $ R_t $, $ U_t $, and $ U_t $ are not necessarily dynamic with time, and it is fairly reasonable to consider these physical parameters as constants for short periods of time. However, in the case of analyzing the system for longer durations, temporal dynamics may apply \cite{delamatar2013downloading}, specifically due to changes in wind speed and velocity. Exact estimation of temporal dynamics related to these parameters are out of the scope of the current study, since we assume locality in time and space according to FARSITE~\cite{finney1998farsite}. The partial derivatives of $ q_t^x $ and $ q_t^y $ with respect to parameters $ R_{t-1} $, $ U_{t-1} $, and $ \theta_{t-1} $ are computed by applying the chain-rule and using Eq.~\ref{eq:qdot1}-\ref{eq:qdot2} as shown in Eq.~\ref{eq:qJacobs1}-\ref{eq:qJacobs2}, where $ \mathcal{L}(\theta) $ equals $ \sin\theta $ and $ \cos\theta $ for X and Y axis, respectively.
	\par\nobreak{\parskip0pt \small \noindent 
		\begin{align}
			\frac{\partial q_{t}}{\partial \theta_{t-1}} &= ~C(R_t, U_t)\frac{\partial\mathcal{L}(\theta)}{\partial\theta}\delta t \label{eq:qJacobs1}\\
			\frac{\partial q_{t}}{\partial R_{t-1}} &= \left(1-\frac{LB(U_t)}{LB(U_t) + \sqrt{GB(U_t)}}\right)\mathcal{L}(\theta)\delta t \\
			\frac{\partial q_{t}}{\partial U_{t-1}} &= \frac{R_{t'}\bigg(LB(U_{t'})\frac{\partial GB(U_{t'})}{\partial U_{t'}} - GB(U_{t'})\frac{\partial LB(U_{t'})}{\partial U_{t'}}\bigg)}{\left(LB(U_{t'})+\sqrt{GB(U_{t'})}\right)^2}\mathcal{L}(\theta)\delta t \label{eq:qJacobs2}
	\end{align}}	
	\begin{figure}[t!]
		\centering
		\includegraphics[scale=.175]{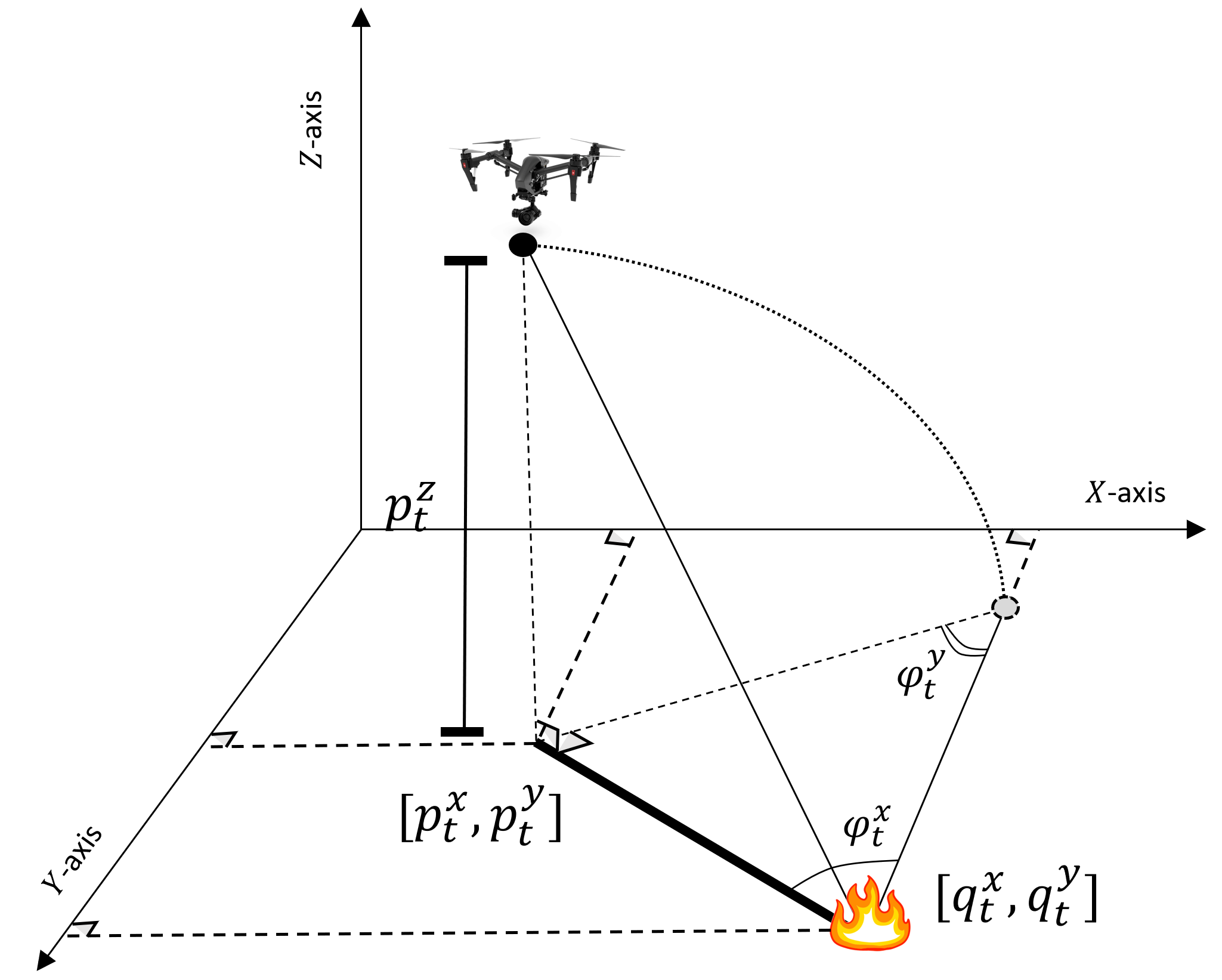}
		\caption{The UAV observation model.}
		\label{fig:ObsMdl}
		\vspace{-0.5cm}
	\end{figure}Next, we derive the observation model through which UAVs perceive dynamic fire-spots, according to Fig.~\ref{fig:ObsMdl}. The observation mapping Eq. in~\ref{eq:probform3} is reformed into Eq.~\ref{eq:probforrm3}
	\par\nobreak{\parskip0pt \small \noindent 
		\begin{equation}
			\label{eq:probforrm3}
			\Bigg[ \hat{\Phi}_t \Bigg]_{5\times1}
			=
			\Bigg[ \left. \frac{\partial h}{\partial \Theta_i}\right|_{\Phi_{t|t}} \Bigg]_{5\times8}
			\Bigg[ \Phi_{t} \Bigg]_{8\times1} + \nu_t, ~\forall i \in \Theta
	\end{equation}}where $ \Phi_t = \left[\varphi_t^x, \varphi_t^y, \hat{R}_t, \hat{U}_t, \hat{\theta}_t\right]^T $ is a mapping vector through which the estimated parameters are translated into a unified, observed \textit{angle}-parameter vector $ \hat{\Phi_{t}} $. The angle parameters (i.e., $ \varphi_t^x $ and $ \varphi_t^y $) contain information regarding both firefront location $ [q_t^x, q_t^y] $ and UAV coordinates $ [p_t^x, p_t^y] $. According to Fig.~\ref{fig:ObsMdl}, by projecting the looking vector of UAV to planar coordinates, the angle parameters are calculated as \nobreak{\parskip0pt \small \noindent$ \varphi_t^x = \tan^{-1}\left(\frac{q_t^x - p_t^x}{p_t^z}\right) $} and \nobreak{\parskip0pt \small \noindent$ \varphi_t^y = \tan^{-1}\left(\frac{q_t^y - p_t^y}{p_t^z}\right) $} for X and Y axes, respectively. Then, the observation Jacobian matrix $ H_t $ is calculated as in Eq.~\ref{eq:observationjacob}
	\par\nobreak{\parskip0pt \scriptsize \noindent
		\begin{multline}
			\label{eq:observationjacob}
			\left. \frac{\partial h}{\partial \Theta_i}\right|_{\hat{\Theta}_{t|t'}} =
			\begin{blockarray}{ccccccccc}
				& q_{t}^x & q_{t}^y & p_{t}^x & p_{t}^y & p_{t}^z & R_{t} & U_{t} & \theta_{t} \\
				\begin{block}{c(cccccccc)}
					\varphi_t^x & \frac{\partial \varphi_{t}^x}{\partial q_t^x} & 0 & \frac{\partial \varphi_{t}^x}{\partial p_t^x} & 0  & \frac{\partial \varphi_{t}^x}{\partial p_t^z} & 0 & 0 & 0 \\	
					\varphi_t^x & 0 & \frac{\partial \varphi_{t}^y}{\partial q_t^y} & 0 & \frac{\partial \varphi_{t}^y}{\partial p_t^y}  & \frac{\partial \varphi_{t}^y}{\partial p_t^y} & 0 & 0 & 0 \\
					\hat{R}_t & 0 & 0 & 0 & 0 & 0 & 1 & 0 & 0 \\
					\hat{U}_t & 0 & 0 & 0& 0 & 0 & 0 & 1 & 0 \\
					\hat{\theta}_t & 0 & 0 & 0 & 0 & 0 & 0 & 0 & 1 \\
				\end{block}
			\end{blockarray}
	\end{multline}}where the partial derivatives are derived as in Eq.~\ref{eq:Hjacobsxaa}-\ref{eq:Hjacobsxx}, using the aforementioned angle parameter equations
	\par\nobreak{\parskip0pt \small \noindent 
		\begin{align}
			\label{eq:Hjacobsxaa}
			\frac{\partial \varphi_{t}^x}{\partial q_t^x} = & \frac{1}{1+\left(\frac{q_t^x-p_t^x}{p_t^z}\right)^2}\left(\frac{1}{p_t^z}\right)\\
			\frac{\partial \varphi_{t}^x}{\partial p_t^x} = & \frac{1}{1+\left(\frac{q_t^x-p_t^x}{p_t^z}\right)^2}\left(\frac{-1}{p_t^z}\right)\\
			\frac{\partial \varphi_{t}^x}{\partial p_t^z} = & \frac{1}{1+\left(\frac{q_t^x-p_t^x}{p_t^z}\right)^2}\left(q_t^x-p_t^z\right)\left(\frac{-1}{(p_t^z)^2}\right) \label{eq:Hjacobsxx}
	\end{align}}similar equations as above hold for \textit{Y}-axis with $ q_t^y $ and $ p_t^y $. 
	
	The process noise $ \omega_t $ in Eq.~\ref{eq:probforrm2} accounts for both stochasticity in fire behavior and wildfire propagation model inaccuracy. Moreover, the observation noise $ \nu_t $ is responsible to account for the estimation errors associated with both $ q_t $ and $ p_t^d $ which affect UAVs’ ability to extrapolate where a fire is on the ground. Taking this into consideration is very important. Both $ \omega_t $ and $ \nu_t $ are modeled as a zero-mean white Gaussian noise with covariances $ Q_t $ and $ \Gamma_t $, respectively. Note that errors in X, Y, and Z axes coordinates of a drone are loosely correlated, and thus, we also incorporate non-diagonal elements in noise covariance matrices when initializing them. $ Q_t $ and $ \Gamma_t $ then receive adaptive updates according to AEKF framework, as in Eq.~\ref{eq:adaptiveQ} and~\ref{eq:adaptiver}.

	\section{Controller Design}
	\label{sec:controllerdesign}
\begin{figure}[t!]
		\centering
		\includegraphics[width=0.9\columnwidth]{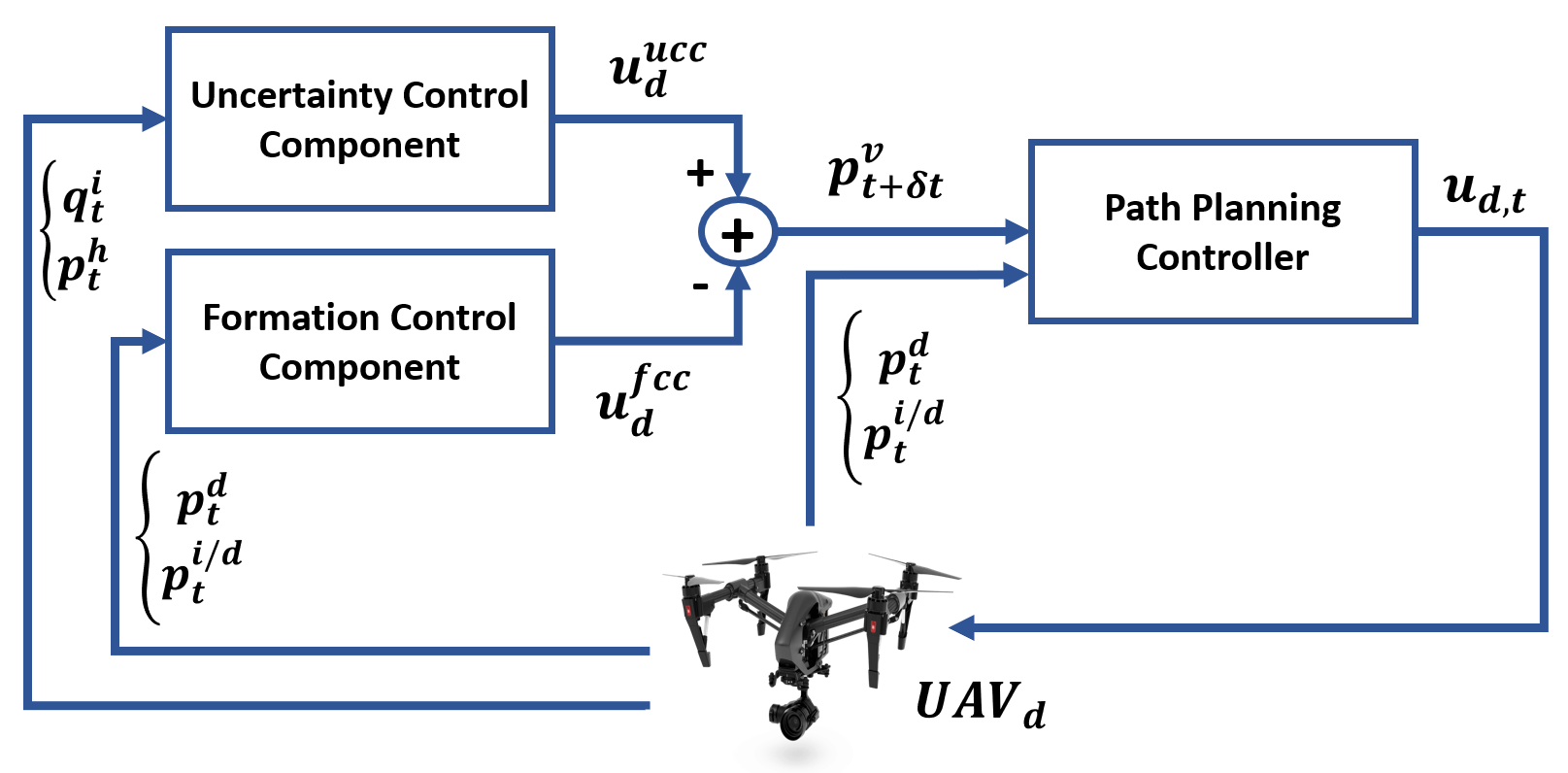}
		\caption{Node-level controller architecture of each drone $ d $.}
		\label{fig:control_architecture}
		\vspace{-0.5cm}
	\end{figure}

	\noindent 
	\noindent Fig.~\ref{fig:control_architecture} represents our node-level controller architecture for each UAV $ d $ with neighboring UAVs $ p_t^{i/d} $. Our controller consists of three components: (1) an uncertainty-based control component (UCC), (2) a formation control component (FCC), and (3) a path planning component (PPC). The first controller performs exploitation to minimize the overall uncertainty in the map as produced (i.e., firefront locations or human areas of activity) while the second controller is designed to manage the general formation of the UAV swarm in order to maximize exploration as well as coverage. The third controller component moves UAVs to any desired next position (i.e., to rendezvous point, to human location for close monitoring, or to next monitoring positions as determined by dual-criteria controller). Assuming $ u_d = \dot{p}_d $ to be the quadcopter UAV dynamics, we develop each of our control components in the following sections.
	
	\subsection{Uncertainty Controller Component (UCC)}
	\label{subsec:UncBasedController}
	\noindent The first controller works based on the theory of artificial potential field~\cite{ge2000new} where each UAV is distributedly controlled by a negative gradient of the generated total uncertainty map $ \mathcal{U}_t^{tot} $ from objective functions in Eq.~\ref{eq:newobjfunc2} and~\ref{eq:disperrorobj}, with respect to its position $ p_t^d=\left[p_t^x, p_t^y, p_t^z\right]^T $ as follows in Eq.~\ref{eq:ctcGD}
	\par\nobreak{\parskip0pt \small \noindent 
		\begin{equation}
			\label{eq:ctcGD}
			u_d^{ucc} = -\kappa_1\frac{\partial\mathcal{U}_t^{tot}}{\partial p_d}
	\end{equation}}where $ \kappa_1 $ is the proportional gain parameter for the first controller. To derive the gradients with respect to the UAV coordinates, we first need to analytically derive the uncertainty objective function (Eq.~\ref{eq:newobjfunc2}). To do so, we insert the values of process and observation Jacobian matrices (i.e., $ F_t $ and $ H_t $) and the process and observation noise covariances (i.e., $ Q_t $ and $ \Gamma_t $) and calculate the trace of the final matrix (see Section~\ref{subsec:OnlineInferenceOfWildfireDynamics}). Eventually, after the simplifications, the CTC objective function equation can be derived as in~\ref{eq:ctcobjfuncequation}
	\par\nobreak{\parskip0pt \small  \noindent
		\begin{align}
			\label{eq:ctcobjfuncequation}
			\min\mathcal{U} &= \underset{q_t^i\in fov_t^d}{\arg\max}\left(\beta_1\left(\frac{\partial\varphi_t^x}{\partial q_t^x}\right)^2 + \beta_2\left(\frac{\partial\varphi_t^y}{\partial q_t^y}\right)^2 + \beta_3\left(\frac{\partial\varphi_t^x}{\partial p_t^x}\right)^2 \right. \nonumber \\
			&~~~~~~+ \left. \beta_4\left(\frac{\partial\varphi_t^y}{\partial p_t^y}\right)^2 + \beta_5\left(\left(\frac{\partial\varphi_t^x}{\partial p_t^z}\right)^2 + \left(\frac{\partial\varphi_t^y}{\partial p_t^z}\right)^2\right) \right)
	\end{align}}where the gradient terms can be calculated using introduced angle-parameters. $ \beta_i $ are covariance constants and are equal to \nobreak{\parskip0pt \small  \noindent$ \beta_1=\left(P_{11}+\sigma_{q^x}^2\right) $, $ \beta_2=\left(P_{22}+\sigma_{q^y}^2\right) $, $ \beta_3=\sigma_{p^x}^2 $, $ \beta_4=\sigma_{p^y}^2 $} and \nobreak{\parskip0pt \small  \noindent$ \beta_5=\sigma_{p^z}^2 $}. Accordingly, the final gradients in Eq.~\ref{eq:ctcGD} with respect to UAV pose can be calculated as in Eq.~\ref{eq:ctcXcontrol}-~\ref{eq:ctcZcontrol}
	\par\nobreak{\parskip0pt \small \noindent 
		\begin{align}
			\frac{\partial\mathcal{U}}{\partial p_d^x} &= \beta_1\frac{\partial\left(\frac{\partial\varphi_t^x}{\partial q_t^x}\right)^2}{\partial p_d^x} + \beta_3\frac{\partial\left(\frac{\partial\varphi_t^x}{\partial p_d^x}\right)^2}{\partial p_d^x} + \beta_5\frac{\partial\left(\frac{\partial\varphi_t^x}{\partial p_d^z}\right)^2}{\partial p_d^x} \label{eq:ctcXcontrol} \\
			\frac{\partial\mathcal{U}}{\partial p_d^y} &= \beta_2\frac{\partial\left(\frac{\partial\varphi_t^y}{\partial q_t^y}\right)^2}{\partial p_d^y} + \beta_4\frac{\partial\left(\frac{\partial\varphi_t^y}{\partial p_d^y}\right)^2}{\partial p_d^x} + \beta_5\frac{\partial\left(\frac{\partial\varphi_t^y}{\partial p_d^z}\right)^2}{\partial p_d^y} \label{eq:ctcYcontrol} \\
			\frac{\partial\mathcal{U}}{\partial p_d^z} &= \beta_1\frac{\partial\left(\frac{\partial\varphi_t^x}{\partial q_t^x}\right)^2}{\partial p_d^z} + \beta_2\frac{\partial\left(\frac{\partial\varphi_t^y}{\partial q_t^y}\right)^2}{\partial p_d^z} + \beta_3\frac{\partial\left(\frac{\partial\varphi_t^x}{\partial p_d^x}\right)^2}{\partial p_d^z} \nonumber \\
			&+ \beta_4\frac{\partial\left(\frac{\partial\varphi_t^y}{\partial p_d^y}\right)^2}{\partial p_d^z} + \beta_5\left(\frac{\partial\left(\frac{\partial\varphi_t^x}{\partial p_d^z}\right)^2}{\partial p_d^z} + \frac{\partial\left(\frac{\partial\varphi_t^y}{\partial p_d^z}\right)^2}{\partial p_d^z}\right) \label{eq:ctcZcontrol}
	\end{align}}and eventually, the control input to UAV $ d $ from UCC module at time $ t $ is noted as in~\ref{eq:ctccontrolinput}
	\par\nobreak{\parskip0pt \small \noindent 
		\begin{equation}
			\label{eq:ctccontrolinput}
			u_{d, t}^{ucc} = \left[\kappa_x\frac{\partial\mathcal{U}_t^{tot}}{\partial p_d^x}, \kappa_y\frac{\partial\mathcal{U}_t^{tot}}{\partial p_d^y}, \kappa_z\frac{\partial\mathcal{U}_t^{tot}}{\partial p_d^z}\right]
	\end{equation}}We note that there is no need to explicitly calculate the gradients of human uncertainty map with respect to UAV positions separately, since we linearly sum up the values (non-negative) of the two maps (see Fig.~\ref{fig:EKF_ErrorMap}).
	
	\begin{figure*}[t!]
		\centering
		\includegraphics[height=5cm, width=1.8\columnwidth]{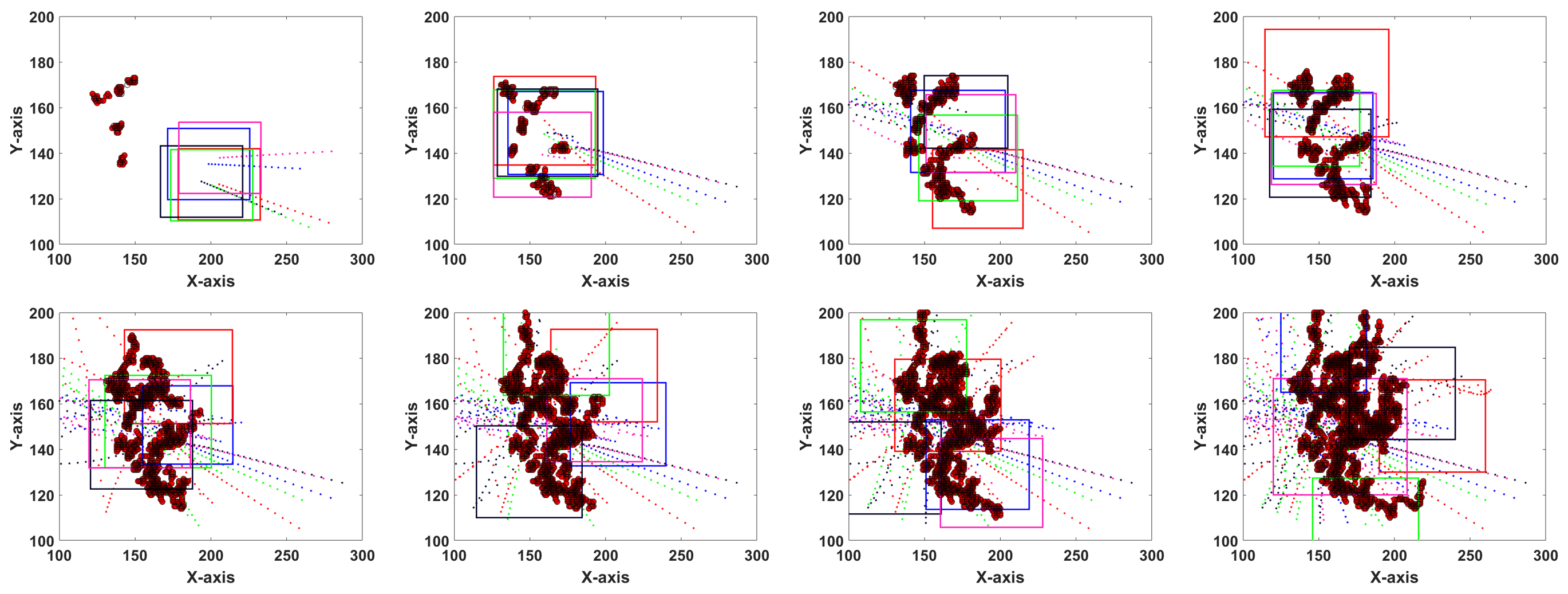}
		\caption{Simulation results of eight sample time-steps between $ t=20 $ (top-left) and $ t=300 $ (bottom-right) for distributed coverage, representing drone FOVs projected on the ground. Dot rays show the FOV centroids.}
		\label{fig:Simul}
		\vspace{-0.5cm}
	\end{figure*}
	\subsection{Formation Controller Component (FCC)}
	\label{subsec:FormationController}
	Similar to the UCC, our formation controller component (FCC) attempts to minimize the consensus displacement error by using a gradient descent flow of the weighted consensus protocol in Eq.~\ref{eq:disperrorobj}, with respect to UAV pose, as represented in Eq.~\ref{eq:fccGD}.
	\par\nobreak{\parskip0pt \small  \noindent
		\begin{equation}
			\label{eq:fccGD}
			u_d^{fcc} = -\kappa_2\frac{\partial\mathcal{E}}{\partial p_d} = -\sum_{(j, d)\in E}\frac{\left(1-\frac{\delta}{\left\|p_t^d-p_t^{d/j}\right\|}\right)\left(p_t^d-p_t^{d/j}\right)}{\left(\Delta-\left\|p_t^d-p_t^{d/j}\right\|\right)^3}
	\end{equation}}Similar logistics as in Eq.~\ref{eq:ctccontrolinput} can be derived here for the three axes of coordinate. Accordingly, the combination of control inputs generated by UCC and FCC modules are leveraged according to our dual-criteria objective function, introduced in Eq.~\ref{eq:dualobjfunc}, in order to produce a new desired location $ p_t^v $ for each UAV to move to. As such, a UAV's new virtual position will be updated and fed to path planning controller (PPC) as shown in Eq.~\ref{eq:newposeupdate}.
	\par\nobreak{\parskip0pt \small \noindent 
		\begin{equation}
			\label{eq:newposeupdate}
			p_{t+\delta t}^v=p_t^v-\left(u_d^{ucc} - u_d^{fcc}\right)\delta t
	\end{equation}}
	\vspace{-1.5em}
	\subsection{Path Planning Controller (PPC)}
	\label{subsec:PPC}
	\noindent The purpose of this controller module is to help a UAV move from its current position to a new position. The PPC module generates either an attractive force toward a desired pose or a repulsive force avoiding an undesirable one. Desired poses include the initial rendezvous point where coverage and tracking wildfire begins and the new virtual position $ p_{t+\delta t}^v $ generated through our dual-criteria objective function as in Eq.~\ref{eq:newposeupdate}. Undesirable poses include ones that are too close to another UAV or too high/low of an altitude such that the drone might  capture low-quality pictures/catch fire. Leveraging an artificial potential field, we address these problems by generating attractive and repulsive forces using a quadratic function of distances from desired or to undesired points. The attractive control force applied to each UAV $ p_t^d $ to any goal points $ p_t^g $ at time $ t $ can be calculated as noted in Eq.~\ref{eq:attforce1}
	\par\nobreak{\parskip0pt \small \noindent 
		\begin{equation}
			\label{eq:attforce1}
			\mathcal{D}_d^{att} = \frac{1}{2}\kappa_g\left\|p_g^d - p_t^d\right\|^2 \ \textrm{and} \ \
			u_d^{att} = -\nabla\mathcal{D}_d^{att} = \kappa_g\left(p_g^d - p_t^d\right)
	\end{equation}}where $ \kappa_g $ is the proportional gain. Using the same notation, the repulsive control force generated to avoid any point $ p_t^g $ can be defined as in Eq.~\ref{eq:repforce1}
	\par\nobreak{\parskip0pt \small \noindent 
		\begin{align}
			\label{eq:repforce1}
			\mathcal{D}_d^{rep} = &
			\begin{dcases}
				\frac{1}{2}\kappa_{g}\left(\frac{1}{\left\|p_g^d - p_t^d\right\|} - \frac{1}{\gamma}\right)^2 & if ~~~\left\|p_g^d - p_t^d\right\| < \gamma \\
				0 & \textrm{otherwise.}
			\end{dcases} \\
			u_d^{rep} = & -\zeta\nabla\mathcal{D}_d^{rep} \\
			= & -\zeta\kappa_{g}\left(\frac{1}{\left\|p_g^d - p_t^d\right\|} - \frac{1}{\gamma}\right)\frac{1}{\left\|p_g^d - p_t^d\right\|^3}\left(p_g^d - p_t^d\right)
	\end{align}}where $ \gamma $ is the distance between current position and the undesirable position $ p_t^g $ and $ \zeta=1 $ only if $ \left\|p_t^g - p_t^d\right\| < \gamma $. Eventually, the general control law in order to generate the required force to move UAVs to their new locations can be formed as in Eq.~\ref{eq:genctrllaw}
	\par\nobreak{\parskip0pt \small \noindent 
		\begin{equation}
			\label{eq:genctrllaw}
			u_{d,t} = \sum_{i}u_{d,t}^{att_i}+u_{d,t}^{rep_i}, ~~\forall i \in F_t
	\end{equation}}where $ F_t $ is the set of all generated attractive and repulsive forces at time $ t $. Thus, the final position of UAV $ d $ gets updated through $ p_{t+\delta t}^d= p_t^d+u_d\delta t $.
	
	\section{Results and Simulation} 
	\label{sec:results}
	\noindent We evaluate the efficiency of our controller in simulation and against two benchmarks: (1) a state-of-the-art, model-based, distributed control algorithm~\cite{pham2017distributed} and (2) a deep reinforcement learning (RL) baseline. The first benchmark~\cite{pham2017distributed} is a fire heat-intensity-based distributed control framework for wildfire coverage which incorporates FARSITE (as in Section~\ref{subsec:simplifiedfarsite}) and a model for fire heat-intensity measure in order to maximize the area-pixel density of the UAV’s fire observations. Furthermore, we train an RL policy network to control UAVs to reduce the uncertainty residual as measured by AEKF. The network consists of four convolutional layers followed by three fully connected layers with ReLU activations. The image of the fire area is an input while a direction for UAVs is an output of the network. We define the reward at each step as the negative sum of uncertainty residual across the entire map, encouraging the agent to minimize uncertainty over time.
	
	In our simulations, we initialize the fire-map with 20 randomly placed ignition points in [50 100] range and within a 500-by-500 terrain where the fire model parameters $ R_t $, $ U_t $, and $ \theta_t $ were chosen similar to~\cite{pham2017distributed}, for comparison. A total of five drones were initialized around [50 300] coordinates with initial altitude set to zero. UAV camera half-angles were set to $ [\frac{\pi}{4}, \frac{\pi}{6}] $. The inter-distance $ \delta $ and communication range $ \Delta $ in our weighted consensus protocol were set to 50 and 500, respectively. The maximum and minimum altitudes were chosen to be $ 15 $ and $ 45 $, respectively. Fig.~\ref{fig:Simul} depicts the simulation results of eight sample time-steps between $ t=20 $ (top-left) and $ t=300 $ (bottom-right) as detailed above, representing drone FOVs projected on the ground.
	
\begin{figure}[t!]
		\centering
		\includegraphics[height=4.5cm, width=\columnwidth]{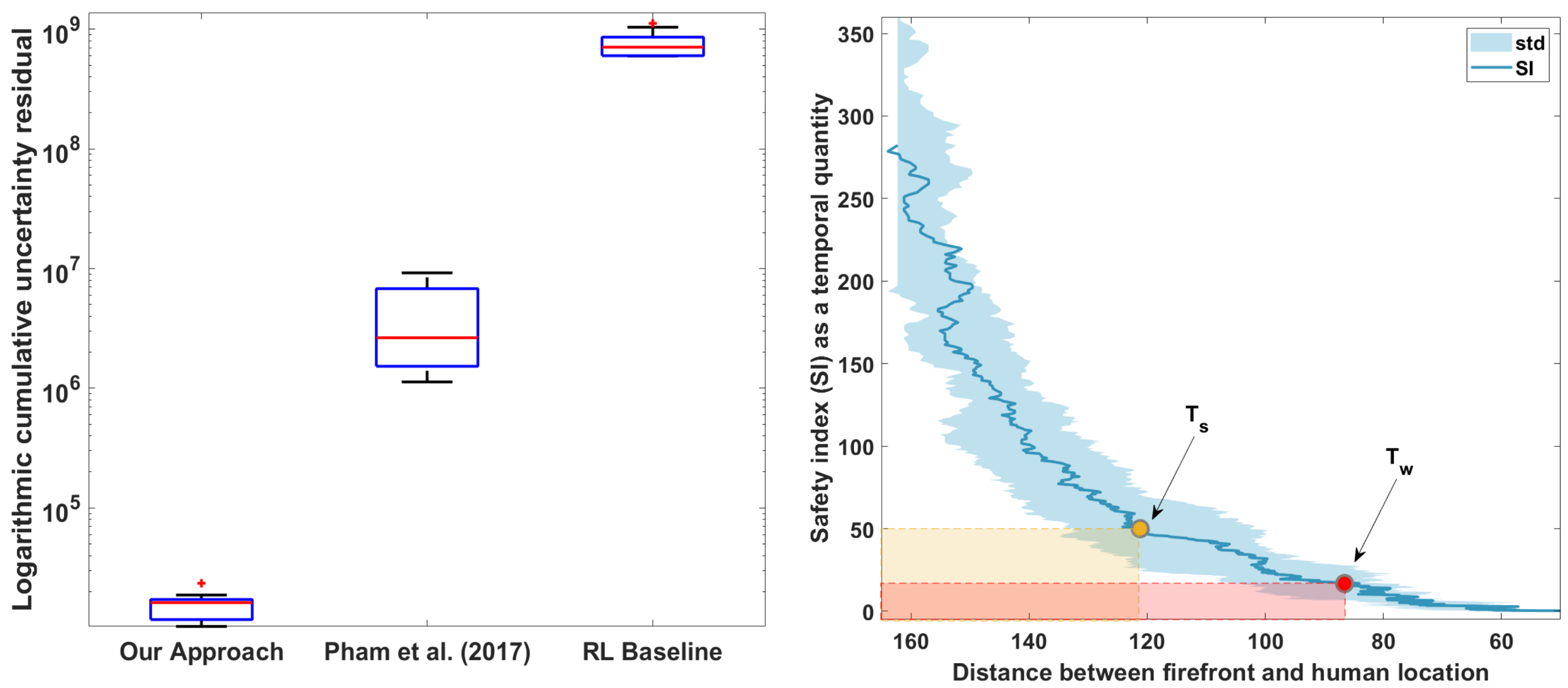}
		\caption{This figure depicts a quantitative comparison of our coordinated controller with prior work (left-side) and human safety index (SI) variations as a temporal quantity, with respect to distance between an approaching firefront and a human firefighter (right-side).}
		\label{fig:Joint_Results_1}
		\vspace{-0.5cm}
	\end{figure}

	The left-side figure in Fig.~\ref{fig:Joint_Results_1} shows a comparison for a team of UAVs controlled by our method, the distributed control proposed by~\cite{pham2017distributed}, and the RL baseline. We ran the simulations for 100 time-steps for all three methods for a total of 10 trials where for each trial, a cumulative uncertainty was calculated by the AEKF for fire points not covered by any drones at each step. While the RL baseline failed to learn during 800 episodes of training, our approach shows significant improvements by reducing the cumulative uncertainty residual by more than $ 10^2 $x and $ 10^5 $x times.
	
\begin{figure}[t!]
		\centering
		\includegraphics[height=4cm, width=\columnwidth]{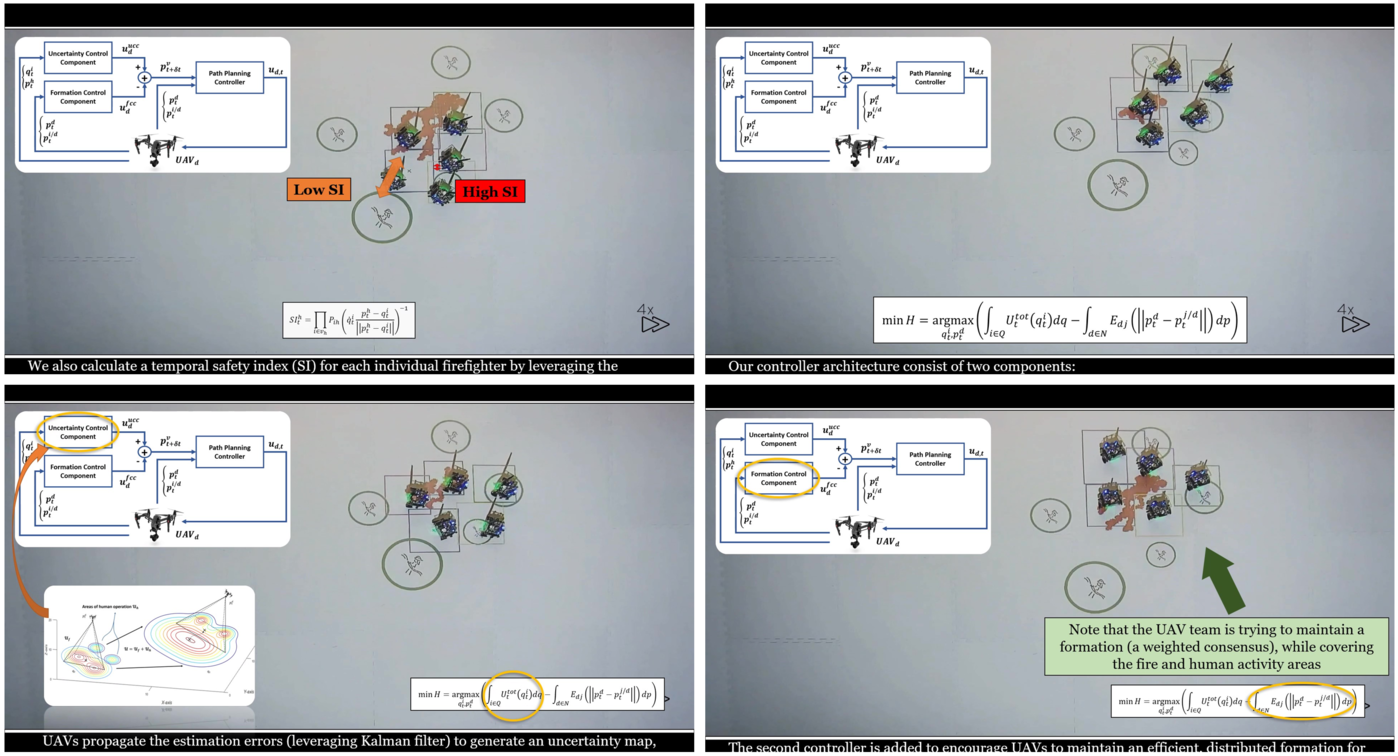}
		\caption{Feasibility of the proposed distributed control algorithm for wildfire coverage evaluated on physical robots in Robotarium platform~\cite{pickem2017robotarium}. The experiment footage can be found at \texttt{https://youtu.be/j3YdIO5u\_fE}.}
		\label{fig:Exp2_split}
		\vspace{-0.5cm}
	\end{figure}

	We also evaluate the feasibility of our controller on physical robots. The physical experiments with actual robots were performed in the Robotarium, a remotely accessible swarm robotics research platform~\cite{pickem2017robotarium}. We tested the coverage performance of our controller using five robots and similar fire environment as above. Fig.~\ref{fig:Exp2_split} represents example demonstrations of our experiment. Results of the experiment is demonstrated in the supplementary video, which can also be found at \texttt{https://youtu.be/j3YdIO5u\_fE}.
	
	\subsection{Safety Index to Secure Human Firefighters}
	\label{subsec:safetyIndex}
	\noindent As a corollary of our algorithm, we calculate an individualized safety index (SI) as a temporal quantity for human firefighters on the ground by leveraging the estimated wildfire dynamics and parameters and report this quantity to firefighters for situational awareness. Now, an individualized safety index (SI) as a measure of time is defined as in Eq.~\ref{eq:safetyindex} for human firefighters by taking into account the velocity of the approaching firefront.
	\par\nobreak{\parskip0pt \small \noindent 
		\begin{equation}
			\label{eq:safetyindex}
			SI_t^h = \prod_{i\in v_h}P_{ih}\left(\dot{q}_t^i\frac{p_t^h-q_t^i}{\left\| p_t^h-q_t^i \right\|}\right)^{-1}
	\end{equation}}In this equation, $ P_{ih} $ is the CDF (from Eq.~\ref{eq:CDFprob}), $ v_h $ is the vicinity of human $ h $, and $ \dot{q}_t^i $ is the estimated fire spread velocity of fire-spot $ i $ toward this vicinity. The ratio is to account for the direction of the firefront and equals to 1 if the firefront is directly approaching the coordinates where the human is located. Accordingly, we assume three different ranges for SI to be announced at each time, namely (1) safe if $ SI_t^h \geq T_s $, (2) warning if $ T_w \leq SI_t^h < T_s $, and (3) danger if $ SI_t^h < T_w $. Parameters $ T_s $ and $ T_w $ are predefined temporal-bounds for safety and warning situations, respectively. We leave the safety and warning thresholds $ T_s $ and $ T_w $ to be pre-defined by humans, as these variables are subjective to the firefighting scenario (e.g., a burning hospital versus forest fire) and are dependent on situational severity. The right-side figure in Fig.~\ref{fig:Joint_Results_1} depicts the variations (i.e. mean$ \pm $std) of SI with respect to distance between an approaching firefront with 10 points and a human firefighter over 100 trials of simulation. For this case, a single UAV was placed over the fire area, inferring the fire-spot locations and parameters.
	
	\section{Conclusion} 
	\label{sec:conclusion}
	\noindent We combined a node-level and an ensemble-level control criteria to introduce a novel coordinated control algorithm for human-centered active sensing of wildfires, providing high-quality, online information to human firefighters on the ground. In our approach, we take advantage of AEKF's error propagation capability to generate a general uncertainty map, incorporating uncertainties about firefront dynamics and areas of human activity. Our approach outperformed prior work for distributed control of UAVs for wildfire tracking as well as a reinforcement learning baseline.
	
	\section*{Acknowledgment}
	\noindent We thank A. Silva for his role in implementing the RL baseline. This work was funded by the Office of Naval Research under grant N00014-19-1-2076. 
	
	\bibliographystyle{IEEEtran}
	\bibliography{IEEEabrv,References_RSS}

\end{document}